\documentclass[letterpaper,11pt,twoside]{article}
\usepackage{fancyheadings}
\usepackage{graphicx}
\usepackage{amsmath}
\def\id{}   % ?? will be replaced with your paper ID number.

\providecommand{\mictitle}[1]{\title{\Large\textbf{#1}}}
\providecommand{\firstaffiliationmark}{$^{\ast}$}
\providecommand{\secondaffiliationmark}{$^{\dagger}$}

\providecommand{\micauthor}[3]{\index{{#2, #1}}#1 #2#3}
\providecommand{\micinstitution}[4]{\normalsize #1#2 \\ #3 \\ \texttt{#4} \\ \hfill \\ }
\providecommand{\institutions}[1]{\date{#1}}
\begin{document}
%%%%%%%%%%%%%%%%%%%%%%%%%%%%%%%%%%%%%%%%%%%%%%%%%%%%%%%%%%%%%%%%%%%%%%%%%%%
%
%      Your final abstract starts here!
%
%%%%%%%%%%%%%%%%%%%%%%%%%%%%%%%%%%%%%%%%%%%%%%%%%%%%%%%%%%%%%%%%%%%%%%%%%%%

% Title of your final abstract
\mictitle{A Greedy Randomized Adaptive Search Procedure for Technicians and Interventions Scheduling for Telecommunications}

% Your name(s): first name, last name and affiliation mark
\author{
        \micauthor{Sylvain}{Boussier}{\secondaffiliationmark}
	      \and
	      \micauthor{Hideki}{Hashimoto}{\firstaffiliationmark}		    
		    \and
		     \micauthor{Michel}{Vasquez}{\secondaffiliationmark}
}

% Your affiliation(s), address(es), and e-mail(s) as they should appear
% in the proceedings volume, preceded by the affiliation mark used for the
% respective author(s) 
\institutions{
%first institution
\micinstitution{\firstaffiliationmark}
   {Graduate School of Informatics, Kyoto University}
    {Kyoto 606-8501, Japan}
    {hasimoto@amp.i.kyoto-u.ac.jp}
%second institution
\micinstitution{\secondaffiliationmark}
    {LGI2P, Ecole des Mines d'Al\`es}
    {Parc Scientifique Georges Besse, F 30035 N\^imes, France}
    {\{Sylvain.Boussier,Michel.Vasquez\}@ema.fr}%third institution
%\micinstitution{\thirdaffiliationmark}
%    {affiliation}
%    {address}
%    {email}
%fourth institution
%\micinstitution{\fourthaffiliationmark}
%    {affiliation}
%    {address}
%    {email}
%fifth institution
%\micinstitution{\fifthaffiliationmark}
%    {affiliation}
%    {address}
%    {email}
%sixth institution
%\micinstitution{\sixthaffiliationmark}
%    {affiliation}
%    {address}
%    {email}
}

\maketitle
\thispagestyle{fancyplain}
\section{Introduction}

The subject of the 5th challenge proposed by the French Society of Operations Research and Decision Analysis (ROADEF)  consists in scheduling technicians and interventions for telecommunications (see \verb+http://www.g-scop.inpg.fr/ChallengeROADEF2007/+ or \verb+http://www.roadef.org/+). The aim of this problem is to assign interventions to teams of technicians. Each intervention has a given priority and the objective is to minimize $z = 28t_1 + 14t_2 + 4t_3 + t_4$ where $t_{\lambda}$ is the ending time of the last scheduled intervention of priority $\lambda$. A schedule is subjected to the following constraints: (1) a team should not change during one day, (2) two interventions assigned on the same day to the same team are done at different times, (3) all predecessors have to be completed before starting an intervention, (4) the working days have a limited duration $H_{max}$ and (5) to work on an intervention, a team has to meet the requirements. Let us note that it is allowed to hire an intervention to an external company but the total amount of cost for hired interventions cannot exceed a given budget $A$. It is easy to show that this problem belongs to the family of NP-hard problems.

\section{Description of the algorithm and notations}
 With experimentation, we noticed that some natural criteria like the coefficient (\emph{linked to the priority of intervention}) in the objective function or even the ratio coefficient/duration are not always the more efficient to insert interventions. The main idea of our approach is to find the best order to insert interventions. The proposed algorithm is divided in three phases: (1) find interventions to be hired and delete them from the problem, (2) find the two \emph{best} orders to insert interventions, (3) generate solutions with a GRASP starting with those two insertion orders.\\
 In what follows, $\Omega_t$ and $\Omega_I$ are respectively the set of technicians and the set of interventions; $R(I,i,n)$  is the number of technicians of level $n$ in domain $i$ required for intervention $I$; $C(t,i)$ is the skill level of technician $t$ in domain $i$; $S(x_i,x_j)=1$ if intervention $i$ is a predecessor of $j$, $0$ otherwise; $T(I)$ is the duration of intervention $I$ and $cost(I)$ is the cost for hiring $I$.
 
 \subsection{Preprocessing heuristics for hired interventions} 
The aim of the preprocessing heuristics is to select a subset of interventions to be hired. The first phase consists in computing a lower bound of the minimum number of technicians needed for each intervention $I$ ($mintec(I)$) by solving the linear program $(P_1(I))$.
 $$
 (P_1(I))
 \begin{cases}
 \text{Minimize } \sum_{t \in \Omega_t} x_t \text{ subject to}, \\
 \sum_{t/C(t,i)\geq n,t \in \Omega_t} x_t \geq R(I,i,n) \hspace{0.5cm} \forall i,n,\\
 x_t \in \{0,1\}\hspace{0.5cm} t \in \Omega_t\\
 \end{cases}
 $$
Let us consider $w_I = mintec(I) \times T(I)$, then we solve the precedence constrained knapsack problem $(KP)$ for each intervention $I$ by considering that the weight of each intervention  is $w_I$. The subset of interventions to be hired is given by the solution of the problem $(KP)$: $I$ is hired if $x_I=1$ and $I$ is not hired otherwise.
 $$
 (KP)
 \begin{cases}
 \text{Maximize } \sum_{I \in \Omega_I} w_I x_I \text{ subject to}, \\
 \sum_{I \in \Omega_I} cost(I) \cdot x_I \leq A,\\
 x_I \leq x_j \hspace{0.5cm} \forall I,j \in \Omega_I / S(x_I,x_j)=1\\
 x_I \in \{0,1\}\hspace{0.5cm} I \in \Omega_I
 \end{cases}
 $$
 
  \subsection{Best insertion orders search phase} 
 
 A priority order $p$ give a weight $\omega_I(p)$ to each intervention $I$. Initially, $\omega_I(p)=28$ for interventions of priority $1$, $14$ for priority $2$, $4$ for priority $3$ and $1$ for priority $4$, in that case $p=(1,2,3,4)$.
 Then we try several runs of a simple greedy algorithm (which insert the interventions with the higher weight first) with the $24$ possible permutations of the $4$ priorities of the problem ($p=(2,3,4,1)$, $p=(3,4,1,2)$, $p=(3,1,4,2)$, etc.) and we keep the two permutations $p_1$ and $p_2$ that give the best \emph{greedy solution}. 
 
 \subsection{Greedy Randomized Adaptive Search Procedure} 

  For each permutation $p=p_1,p_2$, run successively the following algorithm: (1) assign a criteria $C_I = \omega_I(p)$ to each intervention $I$, (2) repeat the following phases until the ending time is reached:
 \begin{itemize}
\item \textbf{Greedy phase}: select the intervention $I$ with the maximum criteria and insert it the earliest day possible, in the team which requires the less additional technicians to perform it and at the minimum starting time possible.
\item \textbf{Local search phase}: if the \emph{greedy algorithm} find a better solution, we try to improve it with \emph{local search}. The \emph{local search} is divided in two phases: (1) the \emph{critical path phase} which search to decrease ending times of priorities ($t_{\lambda}$, $\lambda = 1,2,3,4$) and (2) the \emph{packing phase} which seeks to schedule interventions more efficiently without increasing the ending times of priorities ($t_{\lambda}$).
\item \textbf{Update phase}: Increase the criteria of the last interventions of each priority $I_{\lambda}$ ($\lambda = 1,2,3,4$) and their predecessors $J \in Pred(I_{\lambda})$ so that $C_{I_{\lambda}} := C_{I_{\lambda}} + \omega_{I_\lambda}(p)$ and $C_{J} := C_{J} + \omega_{I_{\lambda}}(p)$.
\end{itemize}

\section{Computational results}
Computational results led us to the 1$^{st}$ position in the Junior category and to the 4$^{th}$ position in All category of the Challenge ROADEF 2007. The results on two data sets provided by France Telecom are exposed in table \ref{tab1}. The description of the data per column is the following:
 \textit{instance}: The name of the instance.
 \textit{int.}: The number of interventions.
 \textit{tec.}: The number of technicians.
 \textit{dom.}: The number of domains.
 \textit{lev.}: The number of levels.
 \textit{best obj.}: The best objective value found by all the challengers.
 \textit{obj.}: The objective value found by our algorithm.
 \textit{gap.}: The gap value to our objective value and the best one.
\begin{table}
\begin{tabular*}{1\textwidth}{@{\extracolsep{\fill}}|c|c|c|c|c|c|c|c|}
\hline
instance&int.&tec.&dom.&lev.&best obj&obj.&gap\\
\hline
\hline
data1-setA&5&5&3&2&2340&2340&0\\
data2-setA&5&5&3&2&4755&4755&0\\
data3-setA&20&7&3&2&11880&11880&0\\
data4-setA&20&7&4&3&13452&13452&0\\
data5-setA&50&10&3&2&28845&28845&0\\
data6-setA&50&10&5&4&18795&18870&0,003\\
data7-setA&100&20&5&4&30540&30840&0,009\\
data8-setA&100&20&5&4&16920&17355&0,025\\
data9-setA&100&20&5&4&27692&27692&0\\
data10-setA&100&15&5&4&38296&40020&0,043\\
%&&&&&&&\\
data1-setB&200&20&4&4&34395&43860&0,215\\
data2-setB&300&30&5&3&15870&20655&0,231\\
data3-setB&400&40&4&4&16020&20565&0,221\\
data4-setB&400&30&40&3&25305&26025&0,027\\
data5-setB&500&50&7&4&89700&120840&0,257\\
data6-setB&500&30&8&3&27615&34215&0,192\\
data7-setB&500&100&10&5&33300&35640&0,065\\
data8-setB&800&150&10&4&33030&33030&0\\
data9-setB&120&60&5&5&28200&29550&0,045\\
data10-setB&120&40&5&5&34680&34920&0,006\\
\hline
\end{tabular*}
\vspace{0.5cm}
\caption{Results obtained on benchmarks provided by France Telecom}\label{tab1}
\end{table}


\begin{thebibliography}{2}   % if you have more than 9 references please use {99}
%
\bibitem{chapter}
Resende MGC, Ribeiro CC. (2003) Greedy randomized adaptive search procedures. In: Glover F, Kochenberger G, editors. {\it Handbook of Metaheuristics}. Dordrecht: Kluwer Academic Publishers, 219–-49.
\end{thebibliography}
\end{document}